\documentstyle[12pt,epsfig]{article}
 1
 1
 1

\begin{document}
\bibliographystyle{unsrt}

\def\lta{\;\raisebox{-.5ex}{\rlap{$\sim$}} \raisebox{.5ex}{$<$}\;}
\def\gta{\;\raisebox{-.5ex}{\rlap{$\sim$}} \raisebox{.5ex}{$>$}\;}
\def\grle{\;\raisebox{-.5ex}{\rlap{$<$}}    \raisebox{.5ex}{$>$}\;}
\def\legr{\;\raisebox{-.5ex}{\rlap{$>$}}    \raisebox{.5ex}{$<$}\;}

\newcommand{\beq}{\begin{equation}} 
\newcommand{\eeq}{\end{equation}}
\newcommand{\bea}{\begin{eqnarray}} 
\newcommand{\eea}{\end{eqnarray}}
\newcommand{\nl}{\nonumber \\} 
\newcommand{\np}{Nucl.\,Phys.\,}
\newcommand{\pl}{Phys.\,Lett.\,}
\newcommand{\pr}{Phys.\,Rev.\,}
\newcommand{\prl}{Phys.\,Rev.\,Lett.\,}
\newcommand{\prep}{Phys.\,Rep.\,}
\newcommand{\zp}{Z.\,Phys.\,}
\newcommand{\sovjnp}{{\em Sov.\ J.\ Nucl.\ Phys.\ }}
\newcommand{\nuclinst}{{\em Nucl.\ Instrum.\ Meth.\ }}
\newcommand{\annp}{{\em Ann.\ Phys.\ }}
\newcommand{\intjmp}{{\em Int.\ J.\ of Mod.\  Phys.\ }}
\newcommand{\ra}{\rightarrow}
\newcommand{\permille}{$^0 \!\!\!\: / \! _{00}$}
\newcommand{\dd}{{\rm d}}
\newcommand{\oal}{{\cal O}(\alpha)}%
\newcommand{\su}{$ SU(2) \times U(1)\,$}
\newcommand{\xdim}{$dim=6$~} 
\newcommand{\eps}{\epsilon}
\newcommand{\mw}{M_{W}}
\newcommand{\mww}{M_{W}^{2}}
\newcommand{\mbb}{m_{b \bar b}}
\newcommand{\mcc}{m_{c \bar c}}
\newcommand{\mbc}{m_{b\bar b(c \bar c)}}
\newcommand{\mh}{m_{H}}
\newcommand{\mhh}{m_{H}^2}
\newcommand{\mz}{M_{Z}}
\newcommand{\mzz}{M_{Z}^{2}}
\newcommand{\lra}{\leftrightarrow}
\newcommand{\tr}{{\rm Tr}}
\newcommand{\ie}{{\em i.e.}}
\newcommand{\cm}{{{\cal M}}}
\newcommand{\cl}{{{\cal L}}}
\def\Ww{{\mbox{\boldmath $W$}}}  
\def\B{{\mbox{\boldmath $B$}}}         
\def\nn{\noindent}
\newcommand{\sinsq}{\sin^2\theta}
\newcommand{\cossq}{\cos^2\theta}
\newcommand{\epem}{$e^{+} e^{-}\;$}
\newcommand{\epemt}{e^{+} e^{-}\;}
\newcommand{\eeah}{$e^{+} e^{-} \ra H \gamma \;$}
\newcommand{\eahnw}{$e\gamma \ra H \nu_e W$}
\newcommand{\thebb}{\theta_{b-beam}}
\newcommand{\thebc}{\theta_{b(c)-beam}}
\newcommand{\pte}{p^e_T}
\newcommand{\ptH}{p^H_T}
\newcommand{\gag}{$\gamma \gamma$ }
\newcommand{\gam}{\gamma \gamma }
\newcommand{\aatoh}{$\gamma \gamma \ra H \;$}
\newcommand{\egam}{$e \gamma \;$}
\newcommand{\eat}{e \gamma \;}
\newcommand{\eaeh}{$e \gamma \ra e H\;$}
\newcommand{\eaehb}{$e \gamma \ra e H \ra e (b \bar b)\;$}
\newcommand{\egebb}{$e \gamma (g) \ra e b \bar b\;$}
\newcommand{\egecc}{$e \gamma (g) \ra e c \bar c\;$}
\newcommand{\egebc}{$e \gamma (g) \ra e b \bar b(e c \bar c)\;$}
\newcommand{\eaebb}{$e \gamma \ra e b \bar b\;$}
\newcommand{\eaecc}{$e \gamma \ra e c \bar c\;$}
\newcommand{\aah}{$\gamma \gamma H\;$}
\newcommand{\zah}{$Z \gamma H\;$}
\newcommand{\zzh}{$Z Z H\;$}
\newcommand{\pe}{P_e}
\newcommand{\pg}{P_{\gamma}}
\newcommand{\delbb}{\Delta m_{b \bar b}}
\newcommand{\delbc}{\Delta m_{b \bar b(c\bar c)}}
\renewcommand{\d}{{\rm d}}
\newcommand{\db}{{\rm d}_{\scriptscriptstyle{\rm B}}}
\newcommand{\bard}{\overline{{\rm d}}}
\newcommand{\bardb}{\overline{{\rm d}}_{\scriptscriptstyle{\rm B}}}
\renewcommand{\aa}{{\rm d}_{\scriptscriptstyle \gamma\gamma}}
\newcommand{\GeV}{\rm GeV}
\newcommand{\TeV}{\rm TeV}
\newcommand{\az}{{\rm d}_{\scriptscriptstyle \gamma Z}}
\newcommand{\zz}{{\rm d}_{\scriptscriptstyle ZZ}}
\newcommand{\aacp}{\overline{{\rm d}}_{\scriptscriptstyle \gamma\gamma}}
\newcommand{\azcp}{\overline{{\rm d}}_{\scriptscriptstyle \gamma Z}}
\newcommand{\zzcp}{\overline{{\rm d}}_{\scriptscriptstyle ZZ}}
\newcommand{\tgw}{\tan{\theta_W}}
\newcommand{\cotgw}{(\tan{\theta_W})^{-1}}
\rightline{LC-TH-1999-015}
\rightline{ROME1-1276/99}
\rightline{December 1999}

\begin{center}
{\Large \bf Anomalous {\boldmath \aah} and {\boldmath \zah} couplings \\
in the process {\boldmath \eaeh}}
\end{center}
\bigskip

\begin{center}
{\large 
E.~Gabrielli~$^{1}$ 
,
$\;\,$V.A.~Ilyin~$^2 \;\,$
and $\;\,$B.~Mele~$^3$. 
} \\
\end{center}
\medskip\noindent
$^1$ Departamento de F\'{\i}sica Te\'orica,
        Universidad Aut\'onoma, Madrid, Spain\\
\noindent
$^2$ Institute of Nuclear Physics, Moscow State University, Russia \\
\noindent
$^3$ INFN, Sezione di Roma 1, and Dipartimento di Fisica, University ``La
Sapienza", Rome, Italy

\bigskip
\begin{center}
{\bf Abstract} \\
\end{center}
{\small 
The bounds on the anomalous contributions to the \aah and \zah vertices that
can be obtained via the process  \eaeh are discussed through  a model
independent analysis based  on \su invariant operators of \xdim. A light Higgs
boson with  $\mh \simeq 120~\GeV$ and center-of-mass energies of
$\sqrt{S}=500~\GeV$ and $\sqrt{S}=1500~\GeV$ are assumed. The process \eaeh is
shown to provide an excellent way to   strongly constraint both the  CP-even
and CP-odd anomalous contributions  in the \aah and \zah couplings.
}

\section{Introduction}

Once the Higgs boson will be discovered, the  next linear \epem colliders with
centre-of-mass (c.m.) energies  $\sqrt{s}\simeq (300\div 2000)$GeV and
integrated luminosity ${\cal O}(100-1000)$ fb$^{-1}$  will allow an accurate
determination of its mass, some couplings and parity properties 
\cite{saar,zerw}.  Among other couplings, the interaction of the Higgs scalar
with the neutral electroweak gauge bosons,  $\gamma$ and $Z$, are particularly
interesting. Indeed, an accurate determination of these couplings  will allow
to test some delicate feature of the Standard Model: the relation between the
spontaneous symmetry breaking mechanism and the electroweak mixing of the two
gauge groups $SU(2)$ and $U(1)$. In this respect, three vertices could be
measured, $ZZH$, $\gamma\gamma H$ and $Z\gamma H$. In the SM the $ZZH$ vertex 
stands at the tree level, while the other two  are generated at one-loop.
Therefore these last ones  are particularly interesting, also because the
effective \aah and \zah couplings could be sensitive to the contributions of
new heavy particles circulating in the loop.

Here, we consider  the case of  a light Higgs boson, that is with $M_Z\lta
\mh\lta 140$ GeV. For this range of mass, a measurement of the \aah coupling
should be possible by the determination of the BR for the decay $H\ra \gamma
\gamma$ in the  channel $gg\to H\to\gamma\gamma$ at the next large hadron
collider LHC (see, for instance, \cite{lhc}).  Furthermore, at future
photon-photon colliders\footnote{Two  further options are presently considered
for a high-energy $e^+e^-$ linear collider,  where  one or both the initial
$e^+/e^-$ beams are replaced by   photon beams induced by Compton
backscattering of laser light on the high-energy electron beams \cite{spec}.
Then, the initial real photons could be to a good degree monochromatic, and
have energy and luminosity comparable to the ones of the parent electron beam
\cite{mono}.}, the precise measurement of the  \aah  vertex can be achieved at
the resonant production of the Higgs particle, $\gamma\gamma\to H$. To this
end, the capability of tuning the \gag c.m.  energy on the Higgs mass, 
through  a good degree of the photons  monochromaticity, will be crucial for
not diluting too much the $\gam \to H$  resonant cross section over the c.m.
energy spectrum. 

Measuring the  \zah vertex is in general a more complicated task. One
possibility to test this vertex is given by  the $H\to \gamma Z$ decay.
Unfortunately, this decay suffers from  having a small rate, and moreover, for
the main $Z$ decays, the signal is affected by large backgrounds.

Another possibility of measuring the \zah vertex  is given by  Higgs production
processes. At electron-positron colliders, the corresponding channels are  $
e^+e^- \to \gamma H$ and $e^+e^- \to Z H$.   The reaction $e^+e^-\to \gamma H$
has been extensively studied in the literature \cite{barr,abba,djou}.
Unfortunately, the \eeah channel suffers from small rates, which are further
depleted at large energies by the $1/s$ behavior of the dominant s-channel
diagrams. The crossed process, the one-loop Higgs production in electron-photon
collisions through  $e\gamma \to eH$, was analysed in \cite{noii,noirep,cott}.
This channel turns out to be a golden place to test  both the \aah and \zah
one-loop  couplings with high statistics, without requiring a fine tuning of
the c.m. energy.  Indeed, the total rate of this reaction is rather high
\cite{noii}.  In particular, for $\mh$ up to about 400 GeV, one finds
$\sigma>1$ fb. If no kinematical cuts are imposed, then the main contribution
to the cross section is given by the \aah vertex. On the contrary, the \zah
vertex contribution  is depleted by the $Z$ propagator.  Nevertheless, the \zah
vertex effects can be  extracted from \eaeh by implementing a suitable strategy
to reduce the \aah vertex contribution \cite{noii}. This strategy requires a
final electron tagged at large angle together with  a transverse momentum cut
$\pte>100$ GeV. The main irreducible background to the process $e\gamma\to
eH\to eb\bar b$ comes from the channel \eaebb.  However, proper cuts on the
quark-pair invariant mass  and on the angles $\thebc$  (between each $b$ quark
direction and both the beams) can bring the signal and background at a
comparable level. Then,  with a luminosity of 100 fb$^{-1}$, at
$\sqrt{s}=500$GeV, one expects an accuracy as good as about 10\% on the
measurement of the \zah effects assuming the validity of the SM \cite{noii}.
\footnote{Since we are discussing the measurement of \aah and \zah couplings 
at the ${\cal O}(10\%)$ level, the QCD corrections to 
the corresponding vertices could be
substantial, and should be taken into account in a precise simulation of the
background in future experiments. Indeed, the next-to-leading QCD corrections 
to the \aah vertex in the SM 
were found to be positive, and at the level of a few per cent for
the Higgs masses discussed here. On the other hand, one can neglect 
the NLO corrections to the \zah vertex. 
One can find the corresponding detailed discussion, e.g., in
\cite{Spira}.}

Here, we present the prospects of the \eaeh reaction in setting experimental
bounds on the value of possible anomalous \aah and \zah couplings \cite{paper}.
Some preliminary results have been presented in \cite{noipri}, too (see also
\cite{banin}). For this analysis we use a model independent approach,  where
\xdim \su invariant operators are added to the SM Lagrangian.  In realistic 
models extending the  SM, these operators contribute in some definite
combinations. However, if one discusses the bounds on possible deviations from 
the SM one-loop Higgs vertices, this approach can give some general insight
into the problem. These anomalous operators contribute to the three vertices
\aah, \zah and $ZZH$, with only the first two involved in the \eaeh reaction.  

We show that the reaction \eaeh has an excellent potential in  bounding both
the vertices \aah and \zah. While in the case of \aah, it is complementary  to
the resonant $\gamma\gamma\to H$ reaction, \eaeh offers a unique possibility to
investigate possible anomalies also in the \zah couplings. 

\section{Anomalous vertices}

We consider the possibility that the new physics affects  the bosonic sector of
the SM through low-energy effective-operators of  \xdim. We restrict our
analysis to the set of operators that contribute to the \eaeh amplitude via
anomalous couplings  to the \aah and \zah vertices. This set can be divided in
two pairs of \xdim  operators\footnote{We assume that the $SU(2)\times U(1)$ 
local gauge invariance of the SM should be valid as well as the so-called
custodial symmetry of the gauge and Higgs sectors that holds in the SM
\cite{anomOP}},  CP-even and CP-odd respectively, giving anomalous
contributions  to the process \eaeh. In terms of these operators  the most
general lagrangian is given by

\beq
{\cal L}^{eff} = \d\cdot {\cal O}_{UW} + \db\cdot {\cal O}_{UB} +
     \bard\cdot \bar {\cal O}_{UW} 
                        + \bardb\cdot \bar {\cal O}_{UB},
\eeq

\beq
{\cal O}_{UW} \,=\, \frac{1}{v^2} \left( |\Phi|^2-\frac{v^2}{2}\right)
    \cdot W^{i\mu\nu}  W^i_{\mu\nu}, \qquad
   {\cal O}_{UB} \,=\, \frac{1}{v^2} \left( |\Phi|^2-\frac{v^2}{2}\right)
    \cdot B^{\mu\nu}  B_{\mu\nu},
\eeq

\beq
\bar {\cal O}_{UW} \,=\, \frac{1}{v^2}  |\Phi|^2
    \cdot W^{i\mu\nu}  \tilde W^i_{\mu\nu}, \qquad
   \bar {\cal O}_{UB} \,=\, \frac{1}{v^2} |\Phi|^2
    \cdot B^{\mu\nu}  \tilde B_{\mu\nu}, 
\eeq
where
$ \tilde W^i_{\mu\nu} = \epsilon_{\mu\nu\mu'\nu'}\cdot W^{i\mu'\nu'}$ and
$ \tilde B_{\mu\nu} = \epsilon_{\mu\nu\mu'\nu'}\cdot B^{\mu'\nu'}$.
In these formulas, $\Phi$ is the Higgs doublet and 
$v$ is the electroweak vacuum expectation value. 
Finally, the \aah and \zah anomalous  contributions to 
the helicity amplitudes of \eaeh are given by:
\beq
M_{anom}(\sigma,\lambda) = 
        M^{\gamma\gamma}(\sigma,\lambda) +
        M^{\gamma Z}(\sigma,\lambda),
\eeq
where
\bea
M^{\gamma\gamma}(\sigma,\lambda) &=&
     \frac{4\pi\alpha}{M_Z (-t)} \sqrt{-\frac{t}{2}}
    \{ d_{\gamma\gamma} [(u-s)-\sigma\lambda(u+s) 
      -i \bar d_{\gamma\gamma} [\lambda (u-s)+\sigma (u+s)]\}, \nl
M^{\gamma Z}(\sigma,\lambda) &=&
     \frac{4\pi\alpha (-g_e^\sigma)}{M_Z (M_Z^2-t)} \sqrt{-\frac{t}{2}}
    \{ d_{\gamma Z} [(u-s)-\sigma\lambda(u+s) 
      -i \bar d_{\gamma Z} [\lambda (u-s)+\sigma (u+s)]\}. \nonumber
\eea  
Here, $s$, $t$ and $u$ are the Mandelstam kinematical variables
(defined as in \cite{noii}), $\sigma/2=\pm 1/2$
and $\lambda=\pm 1$ are the electron and photon  helicities, respectively. 
The $Z$ charge of the electron is denoted as $g_e^\sigma$.
The anomalous couplings $\d, \db$ (CP-even) and  $\bard, \bardb$ (CP-odd)
contribute to the \aah, \zzh, and \zah interactions in the combinations 
\bea
\aa~&=&~ \tgw~\d~+~\cotgw~\db,~~~\az~=~\d~-~\db\nl
\zz~&=&~ \cotgw~\d~+~\tgw~\db,\nl
\aacp~&=&~ \tgw~\bard~+~\cotgw~\bardb,~~~\azcp~=~\bard~-~\bardb\nl
\zzcp~&=&~ \cotgw~\bard~+~\tgw~\bardb,
\label{eq:anomC}
\eea
where $\theta_W$ is the Weinberg angle
(we assume $\sin^2\theta_W=0.2247$)
\footnote{The adopted values for the other physical parameters of the SM 
are described in \cite{noii}.}.

\section{Bounds on anomalous \aah and \zah couplings}

In this section, we present the numerical results 
for the bounds on the anomalous couplings $\d,~\db$ and $\bard,~\bardb$
which are obtained from the \eaeh process.
\footnote{
Most of the results presented in this work were obtained with the help 
of the CompHEP package \cite{comp}.} 
These bounds have been computed by using the
requirement that no deviation from the SM cross section is observed at the 
95\% CL. In particular, we require \cite{paper}:
\beq
N^{\rm anom}(\kappa) < 1.96 \cdot \sqrt{N^{\rm tot}(\kappa)}, \quad 
   \kappa = \d,\db,\bard\;,\bardb\;, 
\eeq
\beq
N^{\rm tot}(\kappa) = {\cal L}_{int} \cdot [\sigma_S(\kappa)+\sigma_B]\;,
  \quad
   N^{\rm anom}(\kappa) = {\cal L}_{int} \cdot 
     [\sigma_S(\kappa)-\sigma_S(0)]\;.
\eeq
where ${\cal L}_{int}$ is the integrated luminosity, 
$N^{\rm tot}$ and $N^{\rm anom}$ denote respectively 
the total number of observed events and
the anomalous number of events deviating from the expected SM predictions
for the signal.
By $\sigma_S(\kappa)$ we mean the cross  section of the signal 
reaction $e\gamma\to eH\to eb\bar b$ with the anomalous contributions.
$\sigma_S(0)$ is the SM cross section.         
By $\sigma_B$, we denote the 
cross section of the  background processes  \eaebb, \eaecc (with 10\%
probability of misidentifying a $c$ quark into a $b$ quark).
For the kinematical cuts applied see \cite{paper}.
\begin{table}[thb]
\begin{center}
\begin{tabular}{|r||c|c|c|c|c|}
\hline
& $\sqrt{S}=500 {\rm GeV}$ & $ \sqrt{S}=500~\GeV$ 
& $\sqrt{S}=1500 {\rm GeV}$ & $ \sqrt{S}=1500~\GeV$ 
\\ \hline 
$\pe=0 $ & $\pte > 0 $ & $ \pte > 100~\GeV$ 
& $\pte > 0 $ & $ \pte > 100~\GeV$ 
\\ \hline 
   \hline $\d~   \times  10^{3}$ &  $(- 0.73,~0.76)$ & 
                                    $(- 1.2,~1.3)$ &
                                    $(-0.24 ,~0.25 )$ & 
                                    $(-0.25 ,~0.25 )$ 
\\ \hline $\db  \times  10^{3}$ &   $(- 0.25,~0.26)$ & 
                                    $(- 0.70,~3.3)$ &
                                    $(-0.10 ,~0.10 )$ & 
                                    $(-0.17 ,~0.21 )$ 
\\ \hline $\bard~\times  10^{3}$ &  $(- 3.2,~3.3)$ & 
                                    $(- 3.6,~3.7)$ &
                                    $(-1.5 ,~1.5 )$ & 
                                    $(-1.2 ,~1.2 )$ 
\\ \hline $\bardb\times 10^{3}$ &   $(- 1.1,~1.1)$ & 
                                    $(- 1.5,~1.5)$ &
                                    $(-5.6 ,~5.6 )$ & 
                                    $(-0.52 ,~0.53 )$
\\ \hline
\hline
$\pe=1 $ & $\pte > 0 $ & $ \pte > 100~\GeV$ 
& $\pte > 0 $ & $ \pte > 100~\GeV$ 
\\ \hline 
   \hline $d~  \times 10^{3}$ &   $(- 0.89,~0.94)$ & 
                                  $(- 10,~25)$ &
                                  $(-0.4 ,~0.4 )$ & 
                                  $(-3.3 ,~17 )$ 
\\ \hline $\db  \times 10^{3}$ &  $(- 0.24,~0.26)$ & 
                                  $(- 0.65,~1.4)$ &
                                  $(-0.11 ,~0.12 )$ & 
                                  $(-0.19 ,~0.49 )$ 
\\ \hline $\bard~\times 10^{3}$ & $(- 3.9,~3.9)$ & 
                                  $(- 15,~15)$ &
                                  $(-2.5 ,~2.7 )$ & 
                                  $(-9.3 ,~8.1 )$
\\ \hline $\bardb\times 10^{3}$ & $(- 0.97,~0.96)$ & 
                                  $(- 0.97,~0.97)$ &
                                  $(-4.5 ,~4.7 )$ & 
                                  $(-0.31 ,~0.31 )$
\\ \hline
\hline
$\pe=-1 $ & $\pte > 0 $ & $ \pte > 100~\GeV$ 
& $\pte > 0 $ & $ \pte > 100~\GeV$ 
\\ \hline 
   \hline $\d~  \times 10^{3}$ &  $(- 0.63,~0.66)$ & 
                                  $(- 0.83,~0.83)$ &
                                  $(-0.18 ,~0.18 )$ & 
                                  $(-0.16 ,~0.17)$ 
\\ \hline $\db  \times 10^{3}$ &  $(- 0.25,~0.26)$ & 
                                  $(- 0.67,~0.66)$ &
                                  $(-0.093 ,~0.094 )$ & 
                                  $(-0.14 ,~0.14 )$ 
\\ \hline $\bard~\times 10^{3}$ & $(- 3.1,~3.2)$ & 
                                  $(- 2.9,~3.1)$ &
                                  $(-1.2 ,~1.2 )$ & 
                                  $(-0.96 ,~0.98 )$
\\ \hline $\bardb\times 10^{3}$ & $(- 1.2,~1.2)$ & 
                                  $(- 2.3,~2.4)$ &
                                  $(-0.69 ,~0.71 )$ & 
                                  $(-0.88 ,~0.90 )$
\\ \hline
\end{tabular}
\caption[]{Bounds on the CP-even $\d,~\db$
and CP-odd $\bard,~\bardb$ anomalous couplings, 
for $\sqrt{S}=500~\GeV$ and $\sqrt{S}=1500~\GeV$, 
$e$-beam polarizations $\pe=0,1,-1$, 
and for $\pte>0$ and $\pte> 100~\GeV$.
In the case of $\sqrt{S}=500~\GeV$ and $\pte> 100~\GeV$, 
a cut on the final electron angle $\theta_e < 90^o$ is applied. }
\end{center}
\end{table}

In Table 1, we report the main results for the bounds on the
anomalous couplings. The bounds are obtained for  $\mh=120~\GeV$, at
$\sqrt{s}=500~\GeV$ (with ${\cal L}_{int}=$100 fb$^{-1}$) and
$\sqrt{s}=1500~\GeV$ (with ${\cal L}_{int}=$1000 fb$^{-1}$),  and for different
electron-beam polarizations,  $P_e=0,1,-1$. We assume that for each bound  the
only contribution to the deviation in the signal is given by the corresponding
anomalous coupling, switching-off the other three anomalous contributions.  A
more complete analysis for the bounds on the  $\d,~\db,~\bard\;,\bardb$  and
$\aa,~\az,~\aacp\;,\azcp$ couplings can be found in \cite{paper},  where  the
correlations between pairs of different anomalous contributions have been
studied, too.

From the results of Table 1, we draw the following conclusions:
\begin{itemize}
\item The strongest bounds on the CP-even couplings at $\sqrt{s}=500$~GeV
 are at the level of
$|\d| \lta 6 \times 10^{-4}$, obtained at $\pe=-1$, 
and $|\db| \lta 2.5 \times 10^{-4}$
(with no cut on $\pte$), not depending on the $e$ polarization.
At $\sqrt{s}=1500$~GeV, one has 
$|\d| \lta 1.7 \times 10^{-4}$, obtained at $\pe=-1$ and $\pte >100$~GeV, 
and $|\db| \lta 1.0 \times 10^{-4}$
(with no cut on $\pte$), not depending on the $e$ polarization.
\item The strongest bounds on the CP-odd couplings at $\sqrt{s}=500$~GeV
are $|\bard| \lta 3 \times 10^{-3}$ and $|\bardb| \lta 1.0 \times  10^{-3}$, 
that are obtained for $\pe=-1$ and
$\pe=1$, respectively. These bounds are quite 
insensitive to the cuts on $\pte$.
At $\sqrt{s}=1500$~GeV, one has $|\bard| \lta 1.0 \times 10^{-3}$ 
for $\pe=-1$, with $\pte>100$~GeV, and
$|\bardb| \lta 3 \times  10^{-4}$, for $\pe=1$, with $\pte>100$~GeV.
\end{itemize}


It is interesting to compare these results with other  bounds obtained in the
literature from different processes at future linear colliders.  In particular,
the processes $e^+ e^- \to HZ$ and $\gamma \gamma \to H$ have been studied for
a \epem collider at $\sqrt{s}=1$~TeV and  with 80 fb$^{-1}$ by Gounaris et al..
From $e^+ e^- \to HZ$, they get  $|\d| \lta 5 \times 10^{-3}$, $|\db| \lta 2.5
\times 10^{-3}$, $|\bard| \lta 5 \times 10^{-3}$ and $|\bardb| \lta 2.5 \times
10^{-3}$  \cite{eeHZ}. The process $\gamma \gamma \to H$ can do a bit better
and reach the values $|\d| \lta 1 \times 10^{-3}$, $|\db| \lta 3 \times
10^{-4}$, $|\bard| \lta 4 \times 10^{-3}$ and $|\bardb| \lta 1.3 \times
10^{-3}$, assuming a particular photon energy spectrum \cite{gouna96}. These
analysis assume a precision of the measured  production rate equal to
$1/\sqrt{N}$ (with $N$ the total number of events), and neglect possible
backgrounds. \noindent In order to set the comparative potential of our process
with respect to these two processes in bounding the parameters  
$\d,~\db,~\bard,~\bardb$,  we assumed $\sqrt{s}=0.9$~TeV and (conservatively) a
luminosity of  25 fb$^{-1}$ in  \eaeh. We then neglected any background, and
assumed a precision  equal to $1/\sqrt{N}$.  In the case $\pe=0$ and $\pte >0$,
we get  $|\d| \lta 5 \times 10^{-4}$, $|\db| \lta 2 \times 10^{-4}$, $|\bard|
\lta 2 \times 10^{-3}$ and $|\bardb| \lta 8 \times 10^{-4}$.

\noindent
This analysis confirms the excellent potential of the process \eaeh.

Following the conventions of reference \cite{9811413},
one can convert our constrains into upper limits for the {\it new physics}
scale $\Lambda$ that can be explored through 
\eaeh with $\sqrt{s}\simeq 1.5$~TeV and 10$^3$ fb$^{-1}$:
\begin{equation}
\begin{array}{rcl}
|\d| \lta 1.7 \times 10^{-4}     
& \to &    |\frac{f_{WW}}{\Lambda^2}| \lta  0.026  ~~ \TeV^{-2} \\\\
|\db| \lta 1.0 \times 10^{-4}     
& \to &    |\frac{f_{BB}}{\Lambda^2}| \lta  0.015  ~~ \TeV^{-2} \\\\
|\bard| \lta 1.0 \times 10^{-3}     
& \to &   |\frac{\bar f_{WW}}{\Lambda^2}| \lta  0.15 ~~  \TeV^{-2} \\\\
|\bardb| \lta 3.0 \times 10^{-4}     
& \to &   |\frac{\bar f_{WW}}{\Lambda^2}| \lta  0.046  ~~  \TeV^{-2} 
\end{array}
\end{equation}

\noindent
For $f_i\sim 1$ one can explore energy scales up to about  6, 8, 2.6 and about
4.5 TeV, respectively.  At $\sqrt{s} \simeq$ 500 GeV and $10^2$ fb$^{-1}$, the 
corresponding constraints on the couplings are a factor 2 or 3 weaker than
above (reflecting into energy scales $\Lambda$ lower by a factor 1.4 or 1.7,
respectively), mainly because of the smaller integrated luminosity assumed.

\newpage
\section*{Acknowledgements}

E.G. acknowledges the financial support of the TMR network project ref.
FMRX-CT96-0090 and partial financial support from the  CICYT project ref.
AEN97-1678. V.I. was partly supported by the joint RFBR-DFG grant 99-02-04011
and the S. Petersburg grant center.


\end{document}